\documentclass[preprint,aps,floatfix,showpacs,a4paper]{revtex4}

\usepackage{amsmath,amssymb,amsfonts,dcolumn,color,graphicx,graphics,latexsym,placeins,epsfig,subfigure,hyperref}

\newcommand{\be}{\begin{equation}}
\newcommand{\ee}{\end{equation}}
\newcommand{\lb}{\left}
\newcommand{\rb}{\right}

\begin{document}

\title{Energetics of a rotating charged black hole in 5-dimensional supergravity} 

\author{Kartik Prabhu}
\affiliation{Department of Physics and Meteorology\\Indian Institute of Technology, Kharagpur, 721302, India} 

\author{Naresh Dadhich}
\email{nkd@iucaa.ernet.in}
\affiliation{Inter-University Centre for Astronomy and Astrophysics\\Pune 411007, India\\}

\begin{abstract}
We investigate the properties of the event horizon and static limit for a charged rotating black hole solution of minimal supergravity theory in $(1+4)$ dimension. Unlike the four-dimensional case, there are in general two rotations, and they couple to both mass and charge. This gives rise to much richer structure to ergosphere leading to energy extraction even for axial fall. Another interesting feature is that the metric in this case is sensitive to the sign of the Maxwell charge. \\
\end{abstract}


\maketitle

\section{Introduction}

Black holes are very interesting gravitational, as well as,  geometric objects. They essentially mark the region where spatial curvature plays the dominant role in the description of gravity. It would therefore be of interest to study them in various settings, including higher dimension. Stability and uniqueness are the remarkable features of black holes which have played an essential role in our understanding of gravity in four dimensions. In view of the current interest primarily inspired by string/M theory and  supergravity or otherwise, it becomes pertinent to study gravity in higher dimensions in general, and black holes in particular. One example of such a remarkable result is the anti-deSitter/conformal field theory correspondence which relates the bulk properties of a five-dimensional supergravity theory to a conformal field theory on its four-dimensional boundary \cite{Maldacena, Gubser, Witten}.\\

Higher dimensional generalizations of Schwarzschild and Reissner-Nordstr\"{o}m black holes were obtained by Tangherlini \cite{Tangherlini}, and of rotating Kerr black hole by Myers and Perry \cite{Myers Perry}. But it is intriguing that a charged rotating black hole does not fall in line in five dimensions. That is, there exists no five-dimensional analogue of Kerr-Newman black hole of Einstein-Maxwell theory. In four dimensions, charge could be put on a rotating black hole simply by adding it into $\Delta = r^2 - 2Mr + a^2 + q^2$, and charge does not enter anywhere else in the metric. Unfortunately and quite strangely, this does not work for five dimensions. Recently, a charged black hole solution in the limit of slow rotation was constructed in \cite{Aliev2} (also see \cite{Aliev3}). Also, charged rotating black hole solutions have been discussed in the context of supergravity and string theory \cite{Cvetic Youm1}, \cite{Cvetic Youm2}, \cite{Cvetic Youm3}. The solution obtained by Chong et al. \cite{Chong Cvetic} of minimal gauged supergravity theory comes closest to Kerr-Newman analogue. It appears that it does not seem to be possible to put pure charge on the rotating black hole; it has always to be accompanied by some other field. \\

We shall study black hole properties of the Chong et al. solution of supergravity Einstein-Maxwell equation \cite{Chong Cvetic}. In Sec. II, we give the metric and electromagnetic potential, which is followed by discussion of event horizon and static limit in Sec. III and of energy extraction by the Penrose process in Sec. IV. In Sec. V, we discuss the most intriguing feature of dependence of the metric and gravitational field on the sign of charge. We conclude with a discussion.  \\

\section{Charged Rotating Black Hole}
 
The general solution for a rotating charged black hole in $(1+4)$ dimensional minimal supergravity with cosmological constant was obtained by Chong et al \cite{Chong Cvetic}. For simplicity, we set the cosmological constant to zero and write the metric and electromagnetic potential as

	\be\label{metric general}
		\begin{split}
ds^2 = & -\frac{\Delta}{\rho^2} ~dT^2 + \frac{\rho^2}{\Delta} ~dr^2 + \rho^2 ~d\theta^2 + \rho^2\sin^2\theta ~d\Phi^2 + \rho^2\cos^2\theta ~d\Psi^2 \\
      &+ \frac{q}{r^2\rho^2} ~dT \lb[2r^2 ~d\nu + q\lb(1-\frac{r^2}{\rho^2} + \frac{2ab}{q} \rb) ~dT \rb] \\
      &+ \frac{\rho^2}{r^2} \lb(b\sin^2\theta ~d\Phi + a\cos^2\theta ~d\Psi \rb)^2 \\
		\end{split}
	\ee

	\be\label{potential general}
		\mathbf{A} = \frac{\sqrt{3}q}{\rho^2}~dT
	\ee

where
	\be\label{parameters general}
		\begin{split}
		dT    & = dt - a\sin^2\theta ~d\phi - b\cos^2\theta ~d\psi \\
		d\nu  & = b\sin^2\theta ~d\phi + a\cos^2\theta ~d\psi \\		
		\rho^2~d\Phi & = a~dt-\lb(r^2+a^2\rb)~d\phi \\
		\rho^2~d\Psi & = b~dt-\lb(r^2+b^2\rb)~d\psi \\
		\Delta  & = \frac{\lb( r^2 + a^2 \rb)\lb( r^2 + b^2 \rb) + q^2 + 2abq}{r^2} - 2 M \\
		\rho^2    & = r^2 + a^2\cos^2\theta + b^2\sin^2\theta.
		\end{split}
	\ee

Here, the angular coordinates range over, $\theta \in \lb[ 0, \pi/2\rb]$ and $\phi , \psi \in \lb[ 0, 2\pi \rb] $.\\

As noted in \cite{Aliev}, the metric and the potential satisfy the Einstein-Maxwell equations with the Chern-Simon term 	
	\begin{subequations}\label{Einstein maxwell CS}
		\begin{align}
 			& {G_i}^k = \frac{1}{2}T^i_k = \frac{1}{2}\lb(F^{im}F_{km} - \frac{1}{4}\delta^i_k F^{mn}F_{mn}\rb) \label{Einstein eqn}\\
			& \nabla_kF^{ik} + \frac{1}{2\sqrt{3}\sqrt{-g}}\epsilon^{iklmn}F_{kl}F_{mn} = 0 \label{Maxwell CS}
		\end{align}
	\end{subequations}

which are obtained from the action 

	\be\label{lagrangian}
 	S = \int d^5x \sqrt{-g} \lb(R - \frac{1}{4}F_{ik}F^{ik} + \frac{1}{12\sqrt{3}}\epsilon^{ijklm}F_{ij}F_{kl}A_m\rb).
	\ee\\

Also note that the metric reduces to that of the Tangherlini solution for $a = b = 0$, and to the Myers-Perry solution for $q = 0$.

The angular momenta \(J\) and charge \(Q\) of the black hole in Eq.(\ref{metric general}) are given by (see \cite{Chong Cvetic})
	\be\label{J phi psi Q general}
		J_\phi = \frac{\pi}{4}\lb( 2aM + bq\rb); \hspace{15pt} J_\psi = \frac{\pi}{4}\lb( 2bM + aq \rb); \hspace{15pt} Q = \frac{\sqrt{3}\pi}{4}q.
	\ee\\

From above expression, it is clear that rotation of a black hole couples to both mass $M$ and charge $q$. This is manifested in the dependence of angular momenta $J_\phi$ and $J_\psi$ on both $M$ and $q$. This dependence on $q$ shows that angular momentum is stored in the electromagnetic field, the feature that was made note of in \cite{Gauntlett Myers Townsend}. We note that the sign of the charge is also relevant here, as it changes angular momentum which could even get reversed, and it makes a nontrivial change in the metric term with the term with $q$ and also in $\Delta$. All of this is quite in contrast to the four-dimensional Kerr-Newman black hole where angular momentum of the hole does not depend upon the charge, and the metric is neutral to its sign. Furthermore, it is interesting to note that the vanishing of one of the angular momenta, $J_\phi$ or $J_\psi$ requires both $a,b \neq 0$. The point to be noted is that rotation here has two aspects, the intrinsic one as $J_\phi$ or $J_\psi$ and the other through the coupling $aq$ or $bq$, as it appears in the metric. This is a remarkable and distinguishing feature of this supergravity black hole, which is not present in the four-dimensional analogue. This is what makes the ergosphere nonvacuous at $\phi$ and $\psi$ axes. \\[10pt]

Now for simplicity, as well as for comparison with the familiar Kerr-Newman solution, we consider only one rotation, i.e. $a \neq b = 0$. In this case, the metric and potential can be simplified to the following form, which closely mimics  Kerr-Newman solution:

	\be\label{metric}
		\begin{split}
ds^2 = & -\frac{\Delta}{\rho^2} ~dT^2 + \frac{\rho^2}{\Delta} ~dr^2 + \rho^2 ~d\theta^2 + \rho^2\sin^2\theta ~d\Phi^2 + r^2\cos^2\theta ~d\psi^2 \\
      &+ \frac{q}{r^2\rho^2} ~dT \lb[2r^2 ~d\nu + q\lb(1-\frac{r^2}{\rho^2} \rb) ~dT \rb]  \\
		\end{split}
	\ee

	\be\label{potential}
		\mathbf{A} = \frac{\sqrt{3}q}{\rho^2}~dT
	\ee

where $dT, d\nu$ etc. as defined in Eq. (\ref{parameters general}) with $b=0$. 
 
The angular momenta \(J\) and charge \(Q\) of the black hole(\ref{metric}) are given by (see \cite{Chong Cvetic})
	\be\label{J phi psi Q}
		J_\phi = \frac{\pi}{2}a M; \hspace{15pt} J_\psi = \frac{\pi}{4}aq; \hspace{15pt} Q = \frac{\sqrt{3}\pi}{4}q.
	\ee
 
It is interesting to note that $J_\phi$ links to mass $M$, while $J_\psi$ links to charge $q$. Even though $b=0$, $J_\psi $ is nonzero due to the coupling $aq$. That is, even when we put $M=0$, rotation does not get switched off because $J_\psi  \neq 0$. This is in contrast with the Kerr or Kerr-Newman black hole where there is only one rotation which is anchored on mass, and hence it goes off along with mass.\\

For comparison, let us recall the Kerr-Newman solution in four-dimensional Einstein-Maxwell theory given by (see \cite{MTW})

	\be\label{metric KN} 
		ds^2 = - \frac{\Delta}{\rho^2}dT^2 + \frac{\rho^2}{\Delta}dr^2 + \rho^2d\theta^2 + \rho^2\sin^2\theta ~d\Phi^2 
	\ee

	\be\label{potential KN}
		\mathbf{A} = \frac{rq}{\rho^2}dT
	\ee

where now the only change is in
	\be\label{parameters KN}
		\Delta = r^2 + a^2 - 2Mr + q^2
	\ee
If in the metrics, (\ref{metric general}, \ref{metric}), the term with the coefficient $q$ was not present, it would have been the expected natural generalization of Kerr-Newman metric in five dimensions. But then it will not be the solution of the Einstein-Maxwell equations. This is the term which comes from the supergravity and Chern-Simon sector. It is this (and also the expression for $\Delta$ with both rotations) which makes the metric depend upon the sign of charge, $q$. It is intriguing that supergravity seems to be crucial for putting charge on a rotating black hole in five dimensions. This is a completely new feature which we do not quite understand.\\[15pt]  

\section{Event Horizon and Static Limit}

The spacetime has a horizon where the $5$ velocity of a corotating observer turns null, or the surface $r = const.$ becomes null. From Eq.(\ref{metric general}), we have \\
		\be
			\Delta = \frac{\left(r^2+a^2\right)\left(r^2+b^2\right)+q^2+2abq}{r^2}-2 M = 0 
		\ee
	
giving the horizon as 	\\
		\be\label{horizon general}
	 	r_\pm = \left[\left(M-\frac{a^2+b^2}{2}\right)\pm\sqrt{\left(M-\frac{a^2+b^2}{2}\right)^2-\left(q+ab\right)^2}\right]^{1/2}.
		\ee\\

The horizon exists if \(a^2+b^2\leq 2M\) and

		\be\label{horizon existence general}
	 	-M+\frac{(a - b)^2}{2}\leq q \leq M - \frac{(a + b)^2}{2}.
		\ee
	
This defines a region in the \(\left(a,b,q\right)\) space, where the metric represents a black hole and not a naked singularity.\\

The static limit is defined where the time-translation Killing vector $\xi_{(t)} \equiv \partial/\partial t$ becomes null (i.e. $g_{tt} = 0$),	
	\be
	 	r^2+a^2\cos^2\theta+b^2\sin^2\theta-2 M + \frac{q^2}{\rho^2}=0
	\ee

giving
	\be\label{staticlimit general}
		 r_{s\pm} = \left[M-a^2\cos^2\theta - b^2\sin^2\theta\pm\sqrt{M^2-q^2}\right]^{1/2}.
	\ee\\

Considering only the \textit{outer} horizon, \(r_{+}\) and static limit, \(r_{s+}\), it can be verified that the static limit always lies outside the horizon. The region between the two is called the \textit{ergosphere}, where timelike geodesics cannot remain static but can remain stationary by corotating with the black hole with the specific frame dragging velocity at the given location in the ergosphere. This is the region of spacetime where timelike particles with negative  angular momentum and/or negative charge relative to the black hole can have negative energy relative to infinity. Then, energy could be extracted from the hole by the well-known Penrose process.

\begin{figure}
		\centering
			
						\subfigure[ q = -0.9]
						{
							\includegraphics[width=0.24\textwidth]{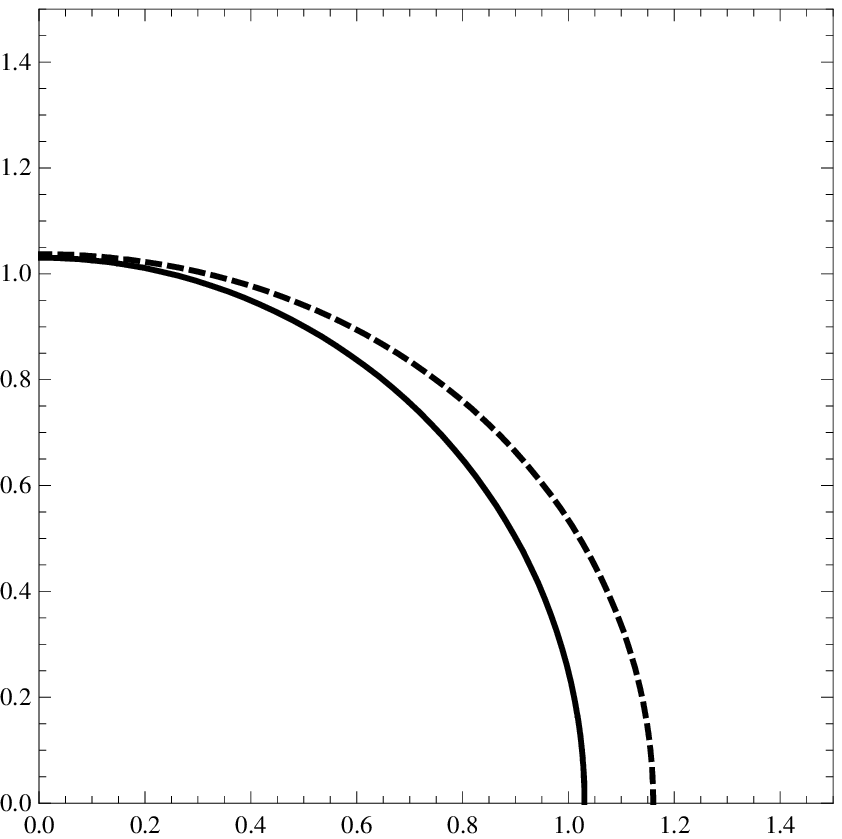} 
							\label{fig b03q-09}
						}
						\subfigure[ q = -0.7]
						{
							\includegraphics[width=0.24\textwidth]{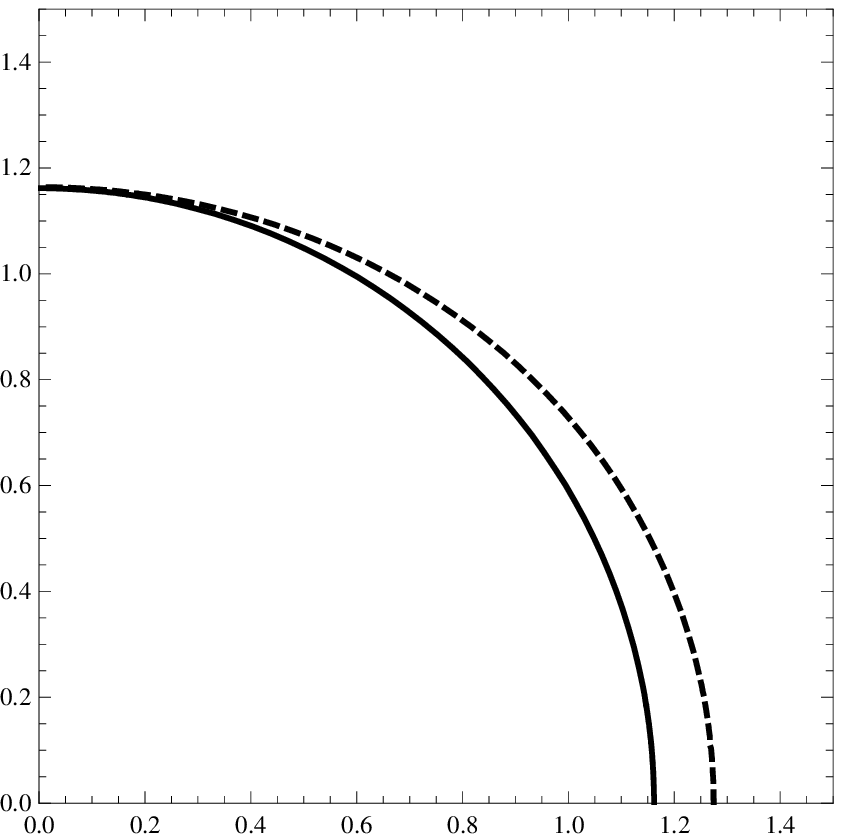} 
							\label{fig b03q-07}
						}
						\subfigure[ q = -0.5]
						{
							\includegraphics[width=0.24\textwidth]{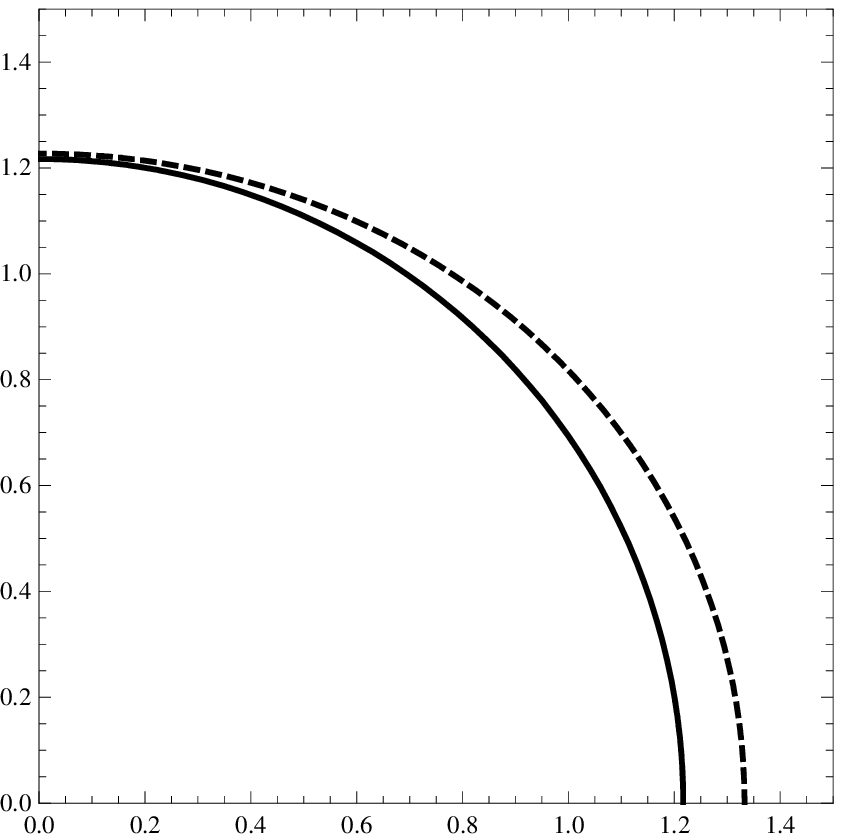} 
							\label{fig b03q-05}
						}
						\subfigure[ q = 0.0]
						{
							\includegraphics[width=0.24\textwidth]{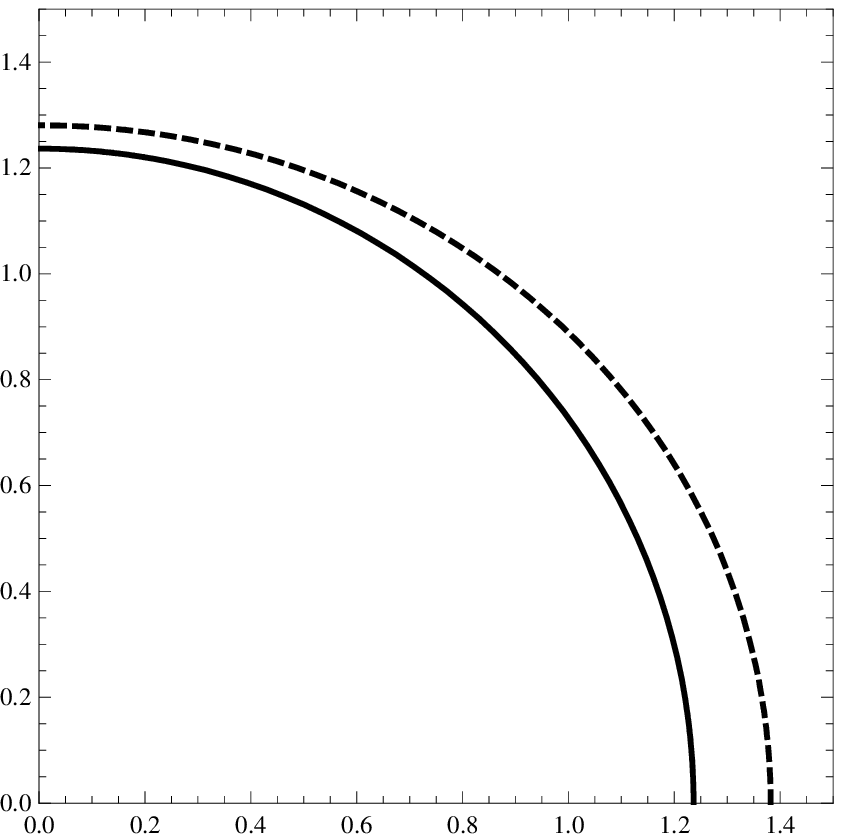} 
							\label{fig b03q00}
						}
						\subfigure[ q = 0.3]
						{
							\includegraphics[width=0.24\textwidth]{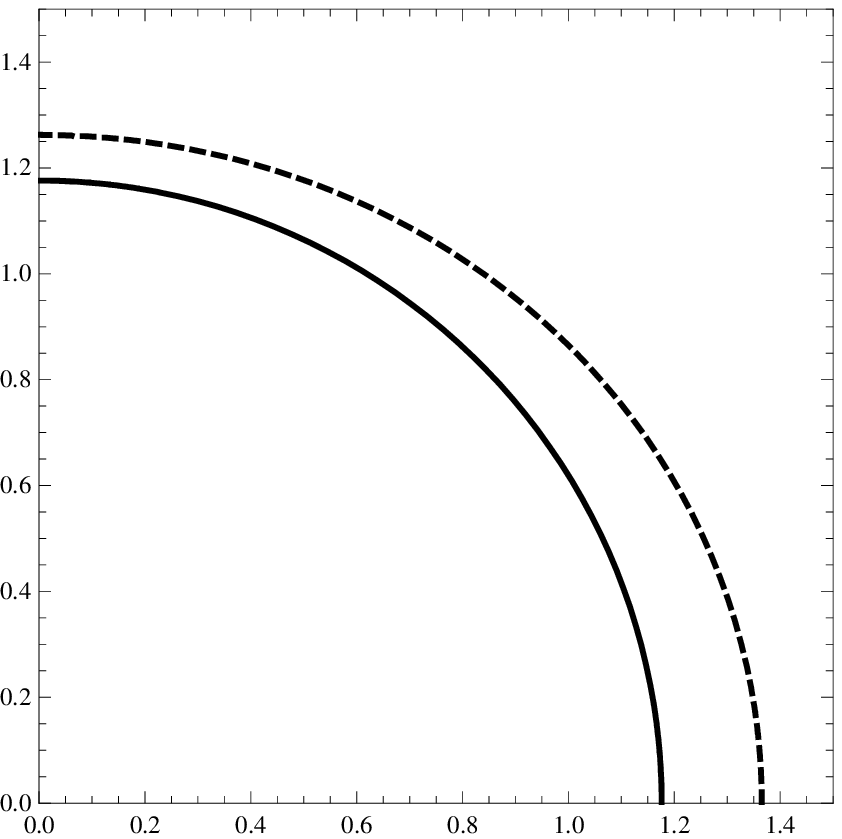} 
							\label{fig b03q03}
						}
						\subfigure[ q = 0.55]
						{
							\includegraphics[width=0.24\textwidth]{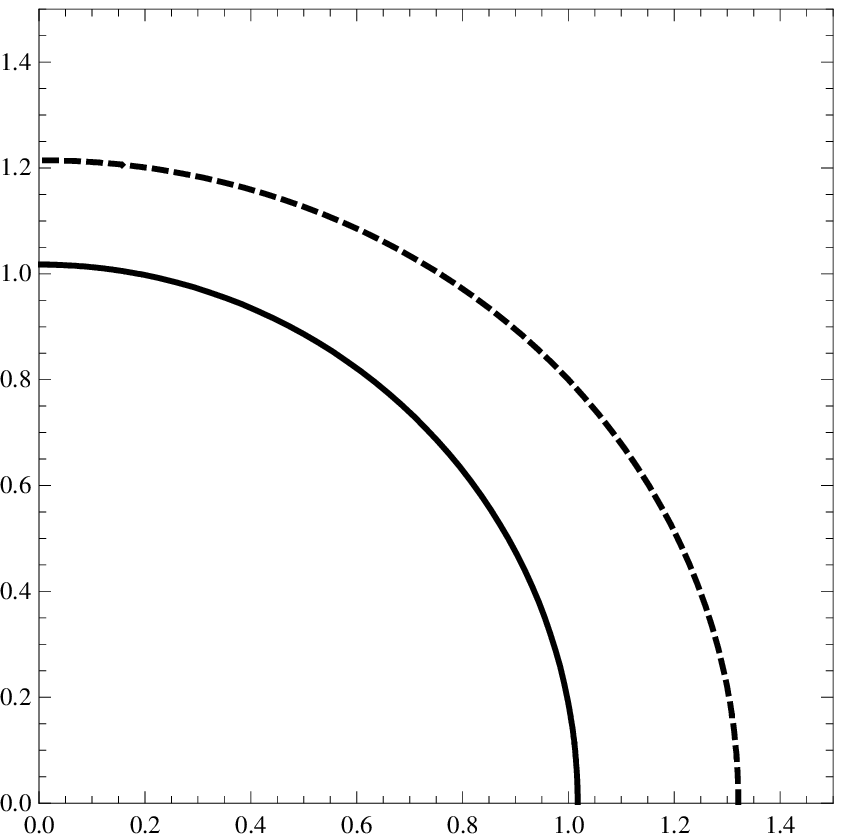} 
							\label{fig b03q055}
						}
						
		\caption{\footnotesize{Polar plots of horizon (solid line) and static limit (dashed line) for \(M = 1.0, a = 0.6, b = 0.3\), and varying \(q\)}}
		\label{horizon varying q}
	\end{figure}

\begin{figure}
		\centering
						\subfigure[ b = -1.0]
						{
							\includegraphics[width=0.24\textwidth]{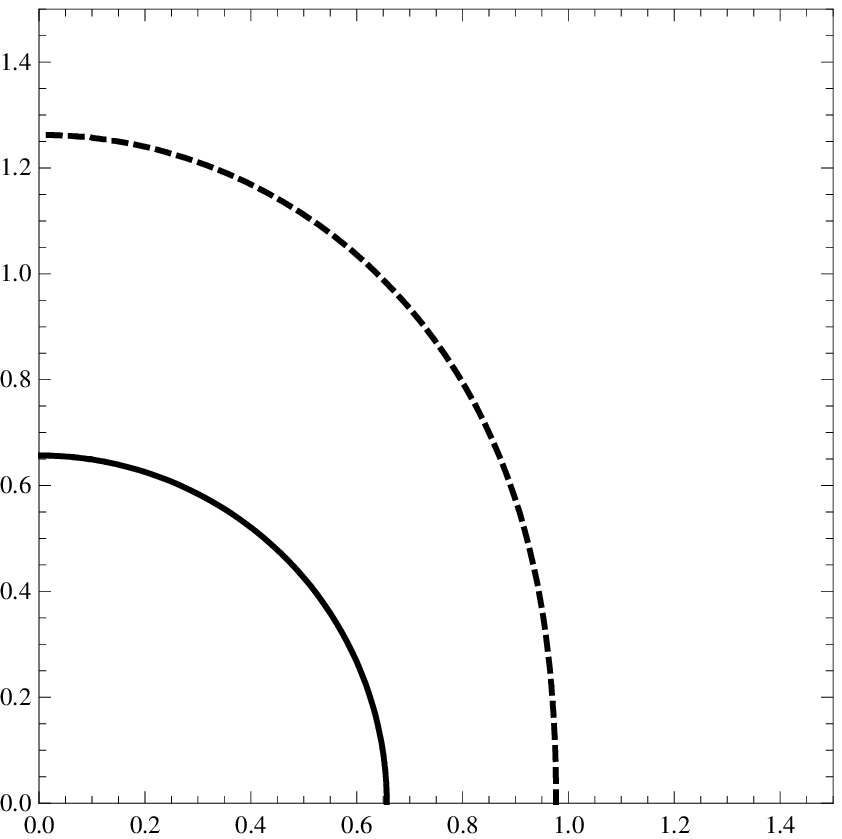} 
							\label{fig q03b-1}
						}
						\subfigure[ b = -0.8]
						{
							\includegraphics[width=0.24\textwidth]{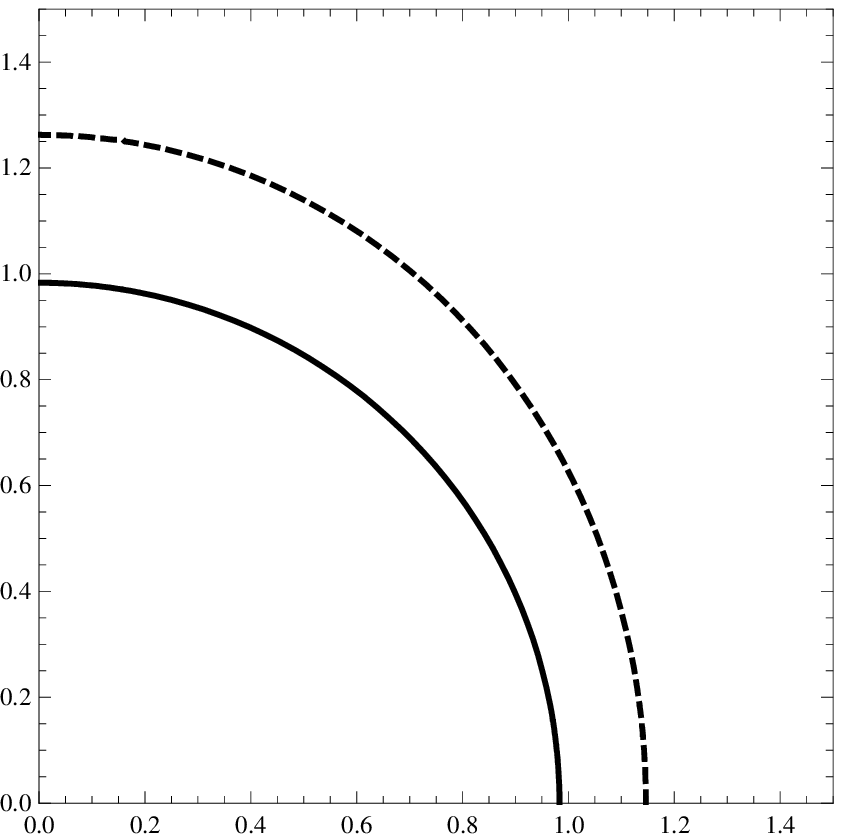} 
							\label{fig q03b-08}
						}
						\subfigure[ b = -0.5]
						{
							\includegraphics[width=0.24\textwidth]{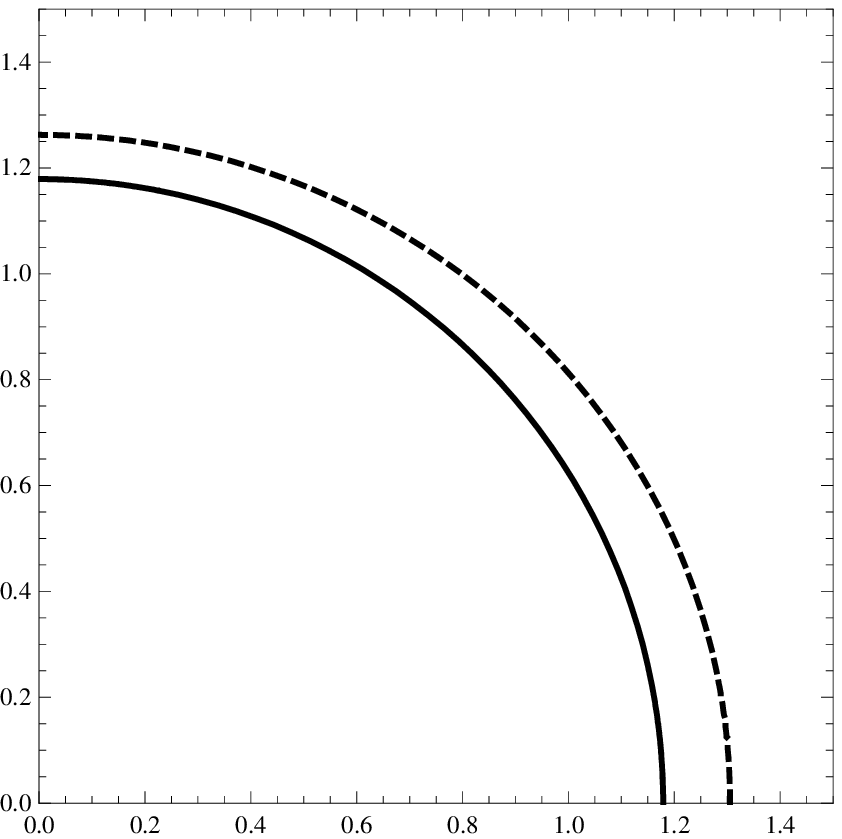} 
							\label{fig q03b-05}
						}
						\subfigure[ b = 0.0]
						{
							\includegraphics[width=0.24\textwidth]{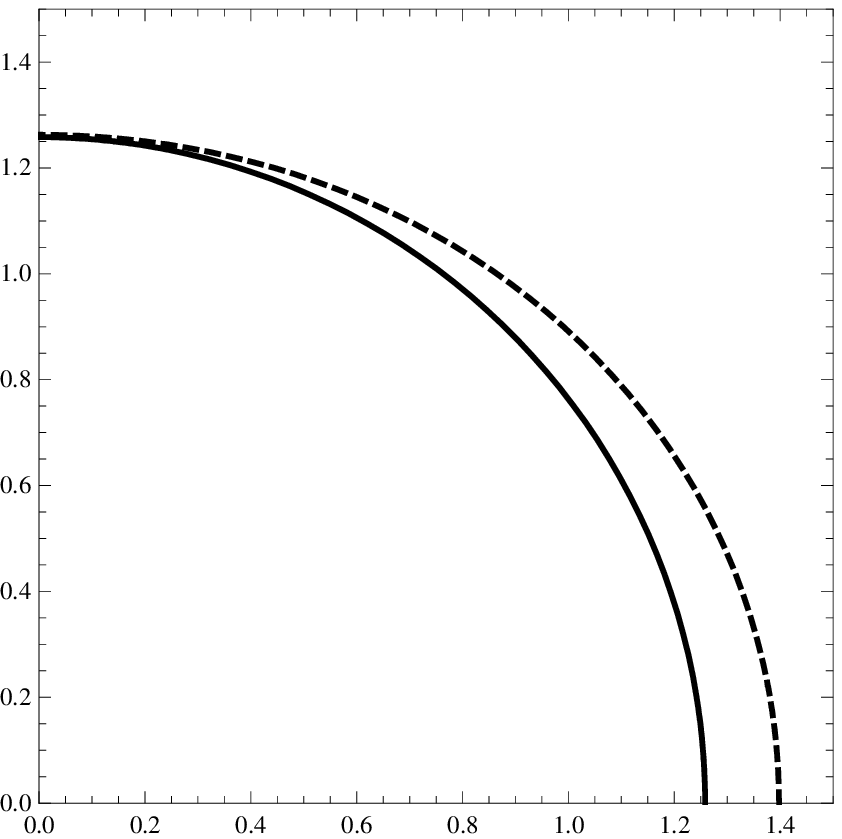} 
							\label{fig q03b00}
						}
						\subfigure[ b = 0.3]
						{
							\includegraphics[width=0.24\textwidth]{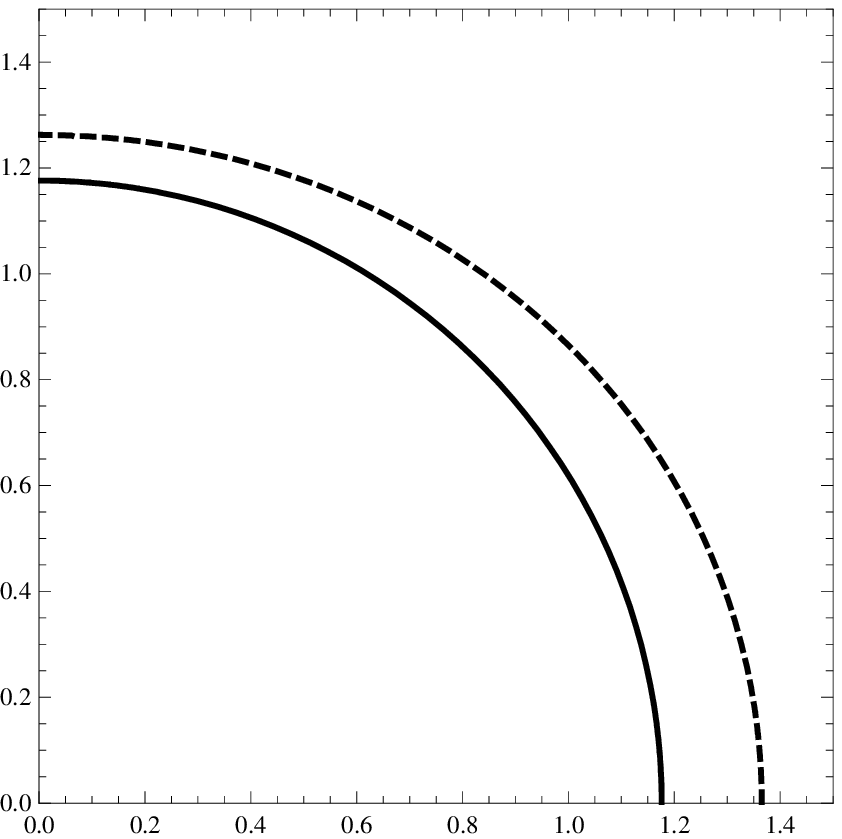} 
							\label{fig q03b03}
						}
						\subfigure[ b = 0.5]
						{
							\includegraphics[width=0.24\textwidth]{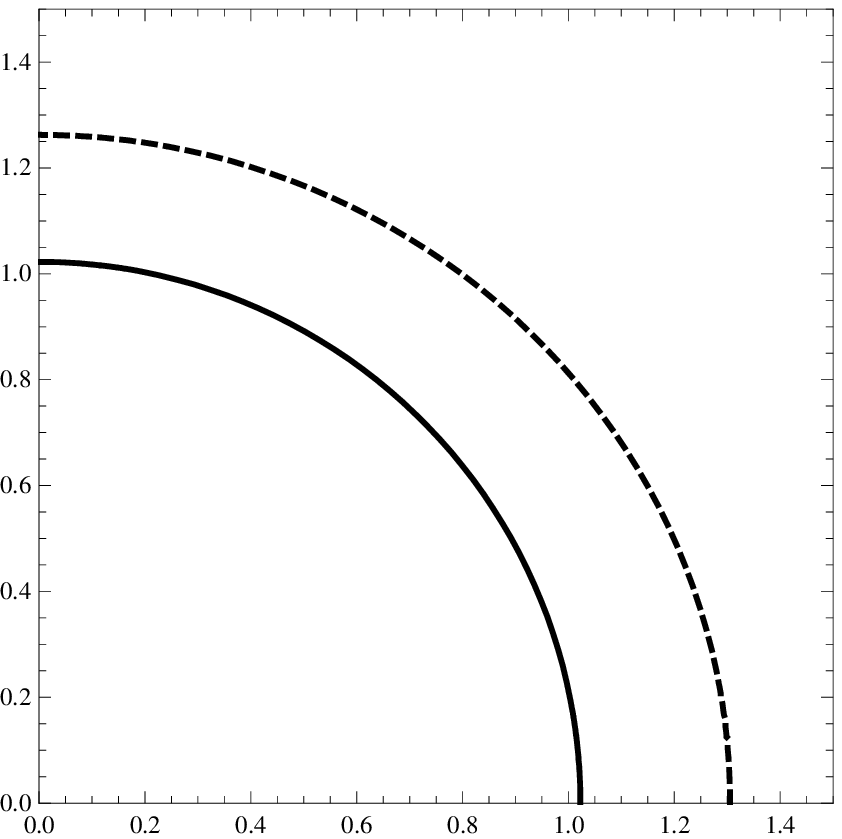} 
							\label{fig q03b05}
						}
						
		\caption{\footnotesize{Polar plots of horizon (solid line) and static limit (dashed line) for \(M = 1.0, a = 0.6, q = 0.3\), and varying \(b\)}}
		\label{horizon varying b}
	\end{figure}

The ergosphere in various cases is shown in Figs.\ref{horizon varying q} and Fig.\ref{horizon varying b} which are polar plots of Eqs.(\ref{horizon general}) and (\ref{staticlimit general}) for fixed $M$ and $a$ with varying $q$ and $b$, respectively.\\

It is clear from Eqs.(\ref{staticlimit general}) and (\ref{horizon general}) that while static limit is insensitive to sign of charge $q$, it is not so for the horizon unless one of the rotations either changes sign or is zero. If there is only one rotation, both the horizon and static limit will be insensitive to the sign of charge $q$, but the metric still depends on the sign of $q$. The dependence of the horizon on the sign of charge could be seen in Figs.\ref{fig b03q-05} and \ref{fig b03q055}. Figure \ref{horizon varying q} shows that for fixed rotation parameters, the separation between the static limit and the horizon starts opening out at $\theta = 0$ with increasing $q$, and it tends to the $\theta=\pi/2$ value at around $q=0.55$. In Fig.\ref{horizon varying b}, we see ergosphere behavior for varying $b$ with all other parameters fixed. Interestingly, here the separation is larger at $\theta=0$ for large negative $b$ which diminishes with increasing $b$, and it tends to the $\theta=\pi/2$ value for $b=0.5$. \\ 
  
Note that the ergosphere exists in general at both $\theta = 0$ and $\theta = \pi/2$. The outer static limit coincides with the horizon (i.e. $r_{s+} = r_+$ ) at $\theta = 0$ if $|a| > |b|$ and at $\theta = \pi/2$ if $|a| < |b|$ when 
		\be\label{coincident q}
			q =-\frac{2abM}{a^2+b^2}.
		\ee\\

\begin{figure}
		\centering
						\subfigure[ b = 0.3]
						{
							\includegraphics[width=0.4\textwidth]{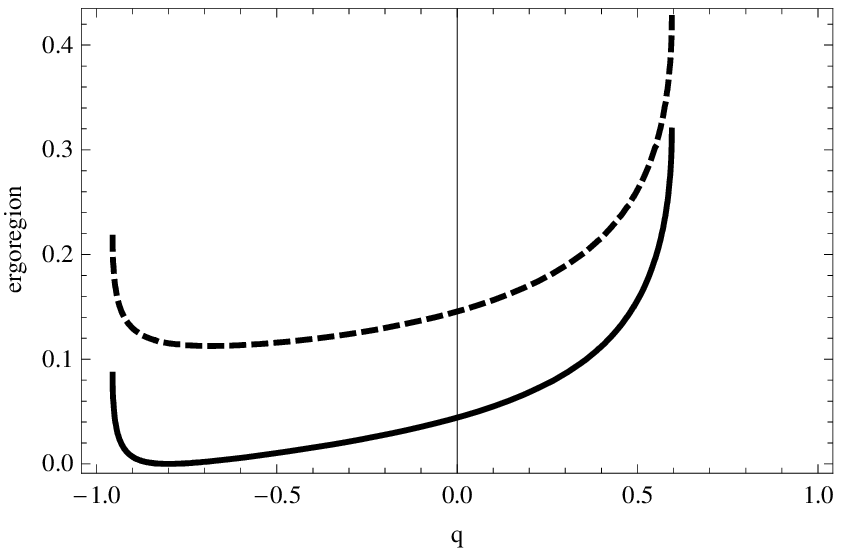} 
							\label{fig ergoregionq}
						}
						\subfigure[ q = 0.3]
						{
							\includegraphics[width=0.4\textwidth]{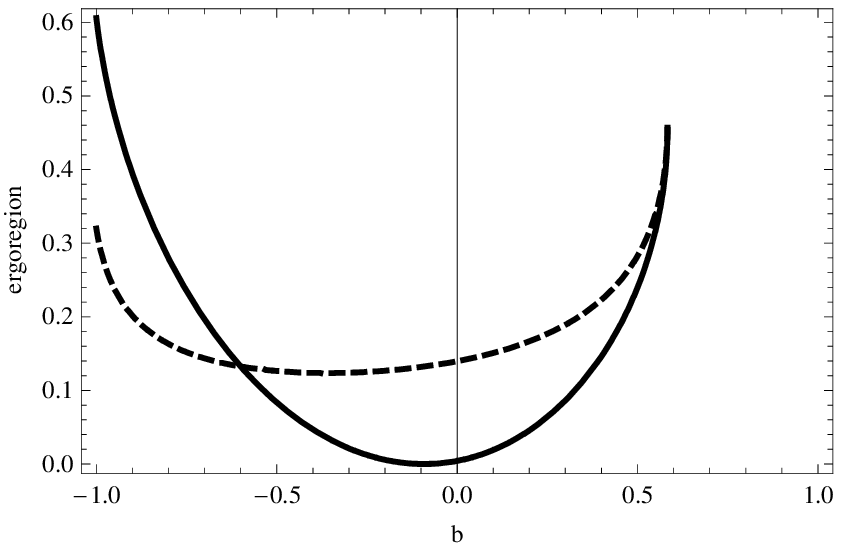} 
							\label{fig ergoregionb}
						}

		\caption{\footnotesize{Plots of ergosphere size at $\theta = 0$(solid) and $\theta = \pi/2$(dashed) for $M = 1.0, a = 0.6$ and varying $q$ and $b$ respectively}}
	\end{figure}

The size of the ergosphere $r_{s+} - r_+$ is plotted in Figs.\ref{fig ergoregionq} and \ref{fig ergoregionb} for varying $q$ and $b$, respectively. These clearly show the peculiar charge dependence of the ergosphere. The plot in Fig.\ref{fig ergoregionq} is not symmetric along the vertical axis, demonstrating the role played by the sign of the charge.

To observe the effect of a single rotation, let us put $b=0$. Now, the horizon and static limits are given by

	\be\label{horizon}
		r_{\pm} = \left[\left(M-\frac{a^2}{2}\right) \pm  \sqrt{\left(M-\frac{a^2}{2}\right)^2 - q^2}\right]^{1/2}.
	\ee

	\be\label{static limit}
		r_{s\pm} = \left[M-a^2\cos^2\theta \pm  \sqrt{M^2 - q^2}\right]^{1/2}.
	\ee	

with

	\be\label{horizon existence}
		\left| q \right| \leq \left|M-\frac{a^2}{2}\right|.
	\ee
	
The distinctive feature of this black hole is the presence of the ergosphere even at the \textit{pole}. At the pole, \(\theta=0\), we have from Eqs.(\ref{horizon}) and (\ref{static limit})
	\[r_+ = \left[\left(M-\frac{a^2}{2}\right) +  \sqrt{\left(M-\frac{a^2}{2}\right)^2 - q^2}\right]^{1/2}\]

and
	\[r_{s+} = \left[M-a^2 +  \sqrt{M^2 - q^2}\right]^{1/2}\].
	
Clearly \(r_+ \leq r_{s+}\) shows the presence of the polar ergosphere. It, however, disappears when \(q = 0\). This happens, because for a rotating black hole, the horizon and static limit coincide at the pole. Obviously, it is relevant only for $\psi$ rotation because there can not be any $\phi$ rotation for polar motion. We shall next show how this ergosphere can be used to extract energy from the black hole for polar accretion. That is, by throwing in particles along $\theta = 0$ with suitable parameters, so they attain negative energy relative to an asymptotic observer.\\[15pt]

\section{Energy Extraction by Penrose Process}

Since there exists an ergosphere around the black hole even at the pole, it is possible to extract energy from it by means of the \textit{Penrose process}. Inside the ergosphere, it is possible to have a timelike or null trajectory with negative total energy. As a result, we can envision a particle falling  from infinity into the ergosphere and splitting into  two fragments, one of which attains negative energy relative to infinity and falls into the hole at the pole, while the other fragment would come out by conservation of energy with energy greater than that of the original incident particle. This is how the energy could be extracted from the hole by axial accretion of particles with suitable angular momentum and charge parameters. Since it is charge which is responsible for the ergosphere at the pole, energy so extracted would be due to electromagnetic interaction and/or interaction between charge and rotation. A similar situation occurs for a Kerr  black hole sitting in a uniform or dipolar magnetic field \cite{dhuran}.\\

Let us now consider the trajectory of such a negative energy particle. The energy and angular momentum of a particle with momentum \(\mathbf{p}\) and electric charge \(e\) are given by 

	\[E = -\mathbf{\pi\cdot \xi_{(t)}}; \hspace{15pt} L_\phi = \mathbf{\pi\cdot \xi_{(\phi)}}; \hspace{15pt}  L_\psi = \mathbf{\pi\cdot \xi_{(\psi)}}\]
		
where $\mathbf{\pi} = \mathbf{p} - e \mathbf{A}$ is the generalized momentum. The $\xi$'s represent the corresponding Killing vectors of the spacetime. Specializing to motion along \(\theta = 0\), and using the metric \ref{metric general} we have \(L_\phi = 0\), \(\rho^2 = r^2+a^2\), and 

	\begin{subequations}\label{E L}
		\begin{align}
			E      & =  \lb(1-\frac{f}{\rho^4}\rb) \pi^t + \lb(\frac{aq}{\rho^2}+\frac{bf}{\rho^4}\rb) \pi^\psi \label{energy}\\
			L_\psi & = -\lb(\frac{aq}{\rho^2}+\frac{bf}{\rho^4}\rb)\pi^t + \lb(r^2+b^2 + \frac{2abq}{\rho^2} + \frac{b^2f}{\rho^4}\rb)\pi^\psi \label{L psi}
		\end{align}
	\end{subequations}\\

where $f = 2 M\rho^2 -q^2$. For timelike trajectories, we have \(-m^2 = g_{ik}p^ip^k\), where \(m\) is the mass of the particle, which gives
	\be
		\begin{split}
			-m^2 = & -\lb(1-\frac{f}{\rho^4}\rb)\left(\pi^t+eA^t\right)^2 + \frac{\rho^2}{\Delta}\left(p^r\right)^2 -2\lb(\frac{aq}{\rho^2}+\frac{bf}{\rho^4}\rb)\left(\pi^t+eA^t\right)\left(\pi^\psi+eA^\psi\right) \\
				   & + \lb(r^2+b^2 + \frac{2abq}{\rho^2}+ \frac{b^2f}{\rho^4}\rb)\left(\pi^\psi+eA^\psi\right)^2	\\
		\end{split}
	\ee\\

Using Eqs.(\ref{energy}) and (\ref{L psi}), this can be written as a quadratic equation in energy:
	\be \label{energy quadratic}
		\alpha E^2 -2\beta E+\gamma + \frac{\rho^2}{\Delta}\left(p^r\right)^2 + m^2 = 0
	\ee

where
	\begin{subequations}\label{alpha beta gamma}
		\begin{align}
		\alpha & = -\lb[r^2+b^2 + \frac{2abq}{\rho^2} + \frac{b^2f}{\rho^4}\rb] D^{-1} \label{alpha}\\
		\beta  & = -\lb[ \frac{L_\psi}{\rho^2} \lb( aq + \frac{bf}{\rho^2} \rb) + 2\sqrt{3}e\frac{q}{\rho^2}\lb( r^2+b^2+\frac{abq}{\rho^2}  \rb) \rb]D^{-1} \label{beta}\\
		\gamma & = \lb[ L_\psi^2\lb( 1- \frac{f}{\rho^4} \rb)- 2\sqrt{3}\frac{L_\psi e q}{\rho^2} \lb( \frac{aq}{\rho^2}+b \rb) - 3 \frac{e^2q^2r^2}{\rho^4} \rb]D^{-1} \label{gamma}\\
		D      & = r^2+b^2 + \frac{2abq}{\rho^2} + \frac{aq^2-fr^2}{\rho^4} \label{D}
		\end{align}
	\end{subequations}

Now, as the particle falls through the horizon, the mass of the black hole will change by \(\delta M = E\). The change in mass can be made as large as one pleases by increasing the mass \(m\) of the infalling particle. But there is a lower limit on \(\delta M\) which could be added to the black hole corresponding to \(m =0\) and \(p^r = 0\) (see \cite{MTW}). Evaluating all of the required quantities at the horizon \(r = r_{+}\), we get the limit for the change in black hole mass as

	\be\label{energy min}
		\delta M \geq E_{min} = \frac{\lb[ a\lb( q + ab \rb) + br_+^2 \rb] L_\psi + \sqrt{3} e ~q r_+^2}{(r_+^2 + a^2)(r_+^2 + b^2) + abq}
	\ee\\

It can be shown that whenever the horizon exists [i.e. Eq.(\ref{horizon existence general}) holds], the denominator is $(r_+^2 + a^2)(r_+^2 + b^2) + abq  > 0$.

We can see the behavior of $E_{min}$, for the limiting cases of single rotation or an uncharged black hole. In these limits, $E_{min}$ reduces to

	\begin{subequations}\label{energy min special}
		\begin{align}
			E_{min}(a=0) & = \frac{bL_\psi + \sqrt{3}eq}{r_+^2 + b^2} \label{energy min a0}\\
			E_{min}(b=0) & = \frac{\frac{aq}{r_+^2}L_\psi + \sqrt{3}eq}{r_+^2 + a^2} \label{energy min b0}\\
			E_{min}(q=0) & = \frac{bL_\psi}{r_+^2 + b^2} \label{energy min q0}
		\end{align}
	\end{subequations}

When the black hole is uncharged (\(q=0\)), energy extraction is possible as long as $b\neq0$. This is the due to the familiar coupling between angular momenta $J_\psi$ and $L_\psi$. But energy extraction is also possible even when $b=0$.  Note that then it is the coupling of charge $q$ with $a$ and $L_\psi$ that is responsible for the energy extraction. This is what is responsible for this interesting and peculiar feature. For the uncharged black hole, there is no energy extraction for axial accretion when $b=0$, as expected.\\

To be able to extract energy from the black hole (\(\delta M < 0\)), we must therefore have

	\[\lb[ a\lb( q + ab \rb) + br_+^2 \rb] L_\psi + \sqrt{3} e ~q r_+^2 < 0.\]

For the special case of $b=0$ using Eq.(\ref{J phi psi Q}), we can write this condition as 

	\be\label{L condition}
		\frac{J_\psi L_\psi}{r_+^2} < - eQ
	\ee

Thus, if \(e\) and \(Q\) are of the same sign, (\(eQ > 0\)), then \(J_\psi\) and \(L_\psi\) must necessarily be of opposite sign. That is, for energy extraction to be possible, the infalling particle must be \emph{counterrotating}, (rotating in the opposite sense) to the black hole's rotation, 
and extracted energy would then come from rotation of the hole.  However, when \(eQ < 0\), energy could still be extracted even for corotation, \(J_\psi L_\psi > 0\), and extracted energy would be purely electromagnetic. In this case, though, there is co-rotation, the particle rotating in the same sense as the hole, yet it  would be lagging behind the corotating frame (i.e. \(\omega_\psi \le \omega_{s\psi}\)), where \(\omega_{s\psi}\) is the angular velocity of the stationary observer with  \(L_\psi = 0\). That is, what really matters for energy extraction is that the particle lags behind the corotating zero angular momentum observer \cite{dhuran}. Also note that even a neutral \(e = 0\) accretion along the axis can extract  energy for the counterrotating particle, which has interestingly been facilitated by the rotation-charge coupling \(aq\).\\

The above analysis can also be carried out for $\theta = \pi/2$ with $a \leftrightarrow b$ and $\phi \leftrightarrow \psi$, and all of the preceding conclusions can be extended to accretion along the \emph{equatorial} region.\\[15pt]

\section{Dependence on sign of charge }
	
We now turn to the strange property of the Chong et. al. black hole, i.e. the metric being sensitive to the sign of charge. If we change only $q$ to $-q$ and keep all other parameters unchanged, the black hole spacetime is also changed and it is different. This is very strange, because charge enters into the field equation through the electromagnetic stress tensor which is quadratic in the electromagnetic field. Hence, it should enter into the metric as $q^2$ which makes the metric  insensitive to its sign. This is the case in general for the Einstein-Maxwell coupling, and in particular, for the four-dimensional Kerr-Newman black hole. It turns out that the Chern-Simon coupling makes one of the stress components, $G^\psi_t = \frac{1}{2}T^\psi_t \sim q^3$, which then becomes sensitive to the sign of $q$. This is what makes the metric sensitive to the sign of charge.\\ 

In the metric Eq.(\ref{metric general}), there are two terms which are sensitive to the sign of $q$: the $q$ term and the term $abq$ in $\Delta$. Both  of these terms couple the charge $q$ to the rotation parameters $a$ and $b$. This makes the angular momenta of the black hole dependent on the sign of $q$ as can be seen in Eq.(\ref{J phi psi Q general}). The most drastic effect can be seen in the special case of $b=0$ from Eq.(\ref{J phi psi Q}). Here, we see that two black holes with opposite charges will have the same $J_\phi$ but opposite $J_\psi$; i.e. the angular momentum along $\psi$ is flipped. Thus the sign of the charge also affects physical quantities that can be observed at infinity.\\

The dependence of $\Delta$ on $q$ makes the horizon dependent on the sign of charge. Equation (\ref{horizon general}) clearly shows that black holes with charges $q$ and $-q$ have different horizons. In fact, from the condition for the existence of the horizon, Eq.(\ref{horizon existence general}), it is clear that by changing the sign of $q$, the black hole can turn into a naked singularity. For instance, for $M = 1.0, a = 0.6, b = 0.3$, it is a black hole for $q = -0.9$ but a naked singularity for $q=0.9$ (Fig.\ref{horizon varying q} shows the effect of charge on the ergosphere.). Thus the existence of the horizon, and hence the black hole, is dependent on the sign of the electric charge. Furthermore it is also reflected in the plot in Fig.\ref{fig ergoregionq} which is not symmetric with the $q$ axis, indicating that the size of the ergosphere is also affected by the sign of the charge. It should however be noted that it is only the horizon that depends on the $q$ sign, while the static limit is insensitive to it. \\

Usually only the motion of charged particles is affected by the sign of the black hole's charge. But since the metric now depends nontrivially on the sign of $q$, the motion of neutral test particles is also affected. The equation of motion for such neutral particles at $\theta = 0$ is described by Eq.(\ref{energy quadratic}) with $e=0$. The effective potential $V(r)$ for the motion of such particles can be readily written using $\lb(p^r\rb)^2 + V(r) = 0$ with 

	\be\label{effective potential}
	V(r) = \frac{\Delta}{\rho^2}\lb(\alpha E^2 -2\beta E+\gamma + m^2 \rb).
	\ee
Since all the parameters $\alpha$, $\beta$, and $\gamma$ depend on the sign of $q$ as can be seen in Eq.(\ref{alpha beta gamma}), even when $e=0$, the sign of the black hole's charge affects the motion of neutral particles. The energetics by the Penrose process will certainly be sensitive to the $q$ sign as can be seen from Eq.(\ref{energy min}), showing the dependence of the energy's lower bound on $q$. The striking case is of $E_{min}$ in Eq.(\ref{energy min b0}) changing sign with $q$ for neutral particles $e=0$ and thereby indicating energy extraction only for $q<0$ and not for $q>0$. \\

\section{Discussion}

One of the puzzling questions is why there does not exist a five-dimensional analogue of the Kerr-Newman solution. The nearest one it has come to is putting a charge on a rotating black hole in gauged supergravity \cite{Chong Cvetic} which is a solution of Einstein-Maxwell-Chern-Simons equations. It seems to raise an interesting question: does charge on a rotating black hole in five dimensions require supergravity? The generalization appears to be on the expected lines but for the new $q$ term and $abq$ term in $\Delta$ involving charge and rotation coupling. It ties the sign of charge with the sense of the hole's rotation. This clearly arises from the Chern-Simons interaction. \\

We have considered energy extraction for the axial accretion (note that there is a symmetry between the two rotations, interchange $a$ and $b$, $\phi$ and $\psi$, and $\theta = 0$ to $\theta = \pi/2$). In particular, energy could be extracted even when $J_\psi = 0$ for the polar $\theta = 0$ case through $aq$ coupling producing a $\psi$ rotation. Here, charge acts as an agent facilitating extraction of rotational energy. This is similar to the \emph{magnetic Penrose process} \cite{wagh} where rotational energy is extracted via electromagnetic interaction. The charge on the black hole can be used to facilitate extraction of rotational energy just as the magnetic field is used in the magnetic Penrose process.\\

One of the distinguishing features of the black hole is that the rotation parameters $a$ and $b$ have a complex relationship with $J_\phi$ and $J_\psi$ through their coupling to charge. That is, $aq$ coupling produces a $J_\psi$ and viceversa. This then gives rise to a much richer ergosphere and to energy extraction scenarios as presented and discussed in Secs III and IV. In particular, rotation anchors on both $M$ and $q$ which means that even when $M =0$, rotation does not get washed out and it presents a nontrivial curved spacetime unlike its Kerr counterpart, flat spacetime for vanishing $M$. Interestingly, rotation about an axis is a composite entity which also has a contribution from the other rotation through its coupling with charge. This is a peculiar feature and is probably due to supergravity. The question arises, is the linkage of rotation to both mass and charge a supergravity property?  It is the latter which causes the nonvacuous ergosphere at the pole allowing for energy extraction for the polar accretion even when $b = 0$.

We have also seen how this solution has a nontrivial dependence on the sign of charge of the black hole, and this manifests itself in various contexts. It is the Maxwell-Chern-Simons coupling which makes the stress tensor $ ~q^3$ and consequently lends the metric sensitive to the sign of $q$. Two black holes with charges $q$ and $-q$ will in general have different angular momenta, horizon structure, ergosphere, and energetics as well as particle orbits. The sign of the charge, thus, affects the spacetime around the black hole. \\

\section{Acknowledgements}

This work arose out of a project KP did under IUCAA's Vacation Students Programme in the summer of 2008 and he thanks IUCAA for hospitality. The authors also thank the anonymous referee for constructive criticism which has resulted in clarifying certain points. \\[20pt]

\end{document}